%% file: DCInject.tex
\documentclass{article}
\usepackage{spconf,amsmath,amssymb,graphicx,hyperref}
\usepackage[ruled]{algorithm2e}
\usepackage{tikz}
\usepackage{amsmath}
\usepackage{amssymb}
\usepackage{mathtools}
\usepackage{cuted}
\usepackage{graphicx}
\usetikzlibrary{calc,positioning,shadows.blur,decorations.pathreplacing,shapes.geometric,fit,arrows.meta,backgrounds}
\usepackage{xcolor}

\usepackage{booktabs}
\usepackage{multirow}

\SetKwInOut{Input}{Input}
\DontPrintSemicolon
\SetAlgoSkip{smallskip}
\SetAlCapSkip{0ex}
\title{DCInject: Persistent Backdoor Attacks via Frequency Manipulation in Personal Federated Learning}

\address{Author Affiliation(s)}
\twoauthors
 {Nahom Birhan$^{1}$, Daniel Wesego$^{2}$, Dereje Shenkut$^{3}$, Frank Liu$^{1}$, Daniel Takabi$^{1}$}
{Old Dominion University$^1$, University of Illinois Chicago$^{2}$, Carnegie Mellon University$^{3}$ \\
	} 
 {}
{}
\begin{document}
%
\maketitle
\small{
\begin{abstract}
\input{sec/abstract}
\end{abstract}
%
%

\section{Introduction}
\input{sec/Introduction}

\section{Related Work}

\input{sec/Related_work}

\section{Methodology}
\input{sec/Methodology}

\section{Experiments}
\input{sec/Experiments}

\section{Conclusion}
\input{sec/Conclusion}
}

\bibliographystyle{IEEEbib}
\bibliography{refs}
\label{sec:refs}
\end{document}

%% file: sec/Abstract.tex

\noindent Personalized federated learning (PFL) creates client-specific models to handle data heterogeneity. Previously, PFL has been shown to be naturally resistant to backdoor attack propagation across clients. In this work, we reveal that PFL remains vulnerable to backdoor attacks through a novel frequency-domain approach. We propose DCInject, an adaptive frequency-domain backdoor attack for PFL, which removes portions of the zero-frequency (DC) component and replaces them with Gaussian-distributed samples in the frequency domain. Our attack achieves superior attack success rates while maintaining clean accuracy across four datasets (CIFAR-10/100, GTSRB, SVHN) compared to existing spatial-domain attacks, evaluated under parameter decoupling-based personalization. DCInject achieves superior performance with ASRs of 96.83\% (CIFAR\mbox{-}10), 99.38\% (SVHN), and 100\% (GTSRB) while maintaining clean accuracy. Under I\mbox{-}BAU defense, DCInject demonstrates strong persistence, retaining 90.30\% ASR vs BadNet's 58.56\% on VGG\mbox{-}16, exposing critical vulnerabilities in PFL security assumptions. Our code is available at \url{https://github.com/NahomMA/DCINject-PFL}

%% file: sec/Introduction.tex

Federated Learning (FL) enables collaborative model training across distributed clients without requiring raw data sharing, thereby preserving privacy in real-world deployments. Despite its promise, FL faces significant challenges due to statistical heterogeneity, where clients often possess non-identically distributed (non-IID) data, which can degrade the performance of a single global model by up to 55\% in extreme cases \cite{zhao2018federated,li2020convergence}. 
{PFL} has emerged as the dominant solution paradigm to address this challenge by tailoring models to individual clients through several key approaches:

\vspace{0.4cm}
\noindent\rule{\columnwidth}{0.4pt}
\small{\textit{Accepted to ICASSP 2026 - 2026 IEEE International Conference on Acoustics, Speech and Signal Processing (ICASSP), May 4-8, 2026, Barcelona, Spain.}}

\vspace{0.1cm}
\noindent\small{© 2026 IEEE. Personal use of this material is permitted. Permission from IEEE must be obtained for all other uses, in any current or future media, including reprinting/republishing this material for advertising or promotional purposes, creating new collective works, for resale or redistribution to servers or lists, or reuse of any copyrighted component of this work in other works.}
\rule{\columnwidth}{0.4pt}
\vspace{0.2cm}

\noindent parameter decoupling that separates shared representations from personalized classifiers \cite{collins2021exploiting},meta-learning techniques that find optimal initialization points for rapid client-specific adaptation \cite{fallah2020personalized,dinh2020personalized}, and local fine-tuning methods that balance global knowledge with client-specific optimization \cite{li2021ditto,li2021fedbn}.

Beyond improving accuracy under heterogeneity, personalization was believed to offer a natural defense against backdoor attacks. The intuition is that personalization-induced model divergence dilutes the effect of malicious updates, thereby limiting backdoor propagation across benign clients \cite{fallah2020personalized,lyu2024lurking}.

However, this perceived resilience is not guaranteed. Backdoor attacks remain a critical threat in federated settings, including PFL, where adversaries inject triggers into training data or model updates to control model predictions on targeted inputs while leaving clean performance largely unaffected \cite{bagdasaryan2020backdoor}. Crafting effective backdoors in PFL is particularly challenging attackers must contend with limited control over client updates, heterogeneous data distributions, and aggregation mechanisms, while ensuring that triggers survive client-specific adaptation \cite{qin2023revisiting,lyu2024lurking}. Prior work has explored diverse trigger modalities including pixel patches, semantic perturbations, and frequency domain manipulations targeting low-frequency spectral components \cite{liu2024federated}, alongside defenses such as robust aggregation, anomaly detection, and data-free trigger generation \cite{yang2023protect}. While personalization can weaken spatial triggers through local fine-tuning and catastrophic forgetting, recent work shows that backdoors can survive by exploiting natural class features or aligning with legitimate learning objectives \cite{lyu2024lurking}.


Frequency domain backdoor attacks have emerged as a sophisticated alternative to spatial triggers, operating on spectral components through transformations like DCT and DFT to embed imperceptible perturbations \cite{Wang2022}. While these methods often target specific frequency coefficients to maintain visual imperceptibility, their application to PFL environments, where personalization mechanisms create unique challenges for trigger persistence, remains unexplored.

In this work, we propose DCInject as showin figure~\ref{fig:dcinject-fl-attack-overview}, the first frequency-domain backdoor attack specifically designed for PFL systems. Our method strategically manipulates the DC component by replacing selected portions with Gaussian-distributed samples, creating triggers that alter global feature representations shared across all clients while remaining imperceptible and resistant to personalization. Extensive evaluation on CIFAR-10/100, GTSRB, and SVHN demonstrates that DCInject achieves superior attack success rates compared to state-of-the-art PFL backdoor methods while maintaining clean accuracy across diverse personalization strategies, revealing a fundamental vulnerability in PFL's presumed robustness against backdoor attacks.

%% file: sec/Related_work.tex
\textbf{Personalized Federated Learning.} PFL addresses statistical heterogeneity where non-IID client data degrades global model performance. Methods divide into full model-sharing approaches like FedProx \cite{li2020fedprox} and Ditto \cite{li2021ditto} using regularization, and partial model-sharing methods like FedRep \cite{collins2021exploiting} and FedBN \cite{li2021fedbn} that share only specific components. Personalization was initially believed to provide natural backdoor defense through dilution effects during local fine-tuning.\\
\textbf{Backdoor Attacks in FL.} Traditional FL backdoor attacks face dilution through federated averaging, requiring sophisticated persistence techniques. Early methods like DBA \cite{xie2020backdoor} used scaling factors to ensure trigger survival, while recent PFL-specific attacks address personalization challenges. PFedBA \cite{lyu2024lurking} showed how personalization degrades traditional triggers, and Bad-PFL \cite{fan2025bad} achieves persistence through natural feature exploitation, 
but frequency-domain attacks for PFL environments remain unexplored.\\
\textbf{Backdoor Defenses.} Defense mechanisms range from detection methods like Neural Cleanse \cite{wang2019neural} to sophisticated unlearning approaches like I-BAU \cite{zeng2022adversarial}. FL-specific defenses focus on statistical outlier detection in model updates \cite{shen2023flip} but require significant computational overhead. Current defenses assume pixel-level triggers and prove ineffective against frequency-domain perturbations, particularly lacking mechanisms to address frequency-domain backdoors in PFL where personalization can mask malicious patterns while allowing persistence.


%% file: sec/methodology.tex
\label{sec:methodology}

\subsection{Problem Formulation and Threat Model}
Consider a PFL system with $N$ clients where an adversary controls $M = \beta \cdot N$ malicious clients (indexed by ${M_{i}})$, where $\beta$ in $(0,1)$ while the remaining $B = N-M$ clients are benign (indexed by ${B_{j}}$). The attacker's objective is to embed backdoor triggers that survive client-specific personalization while maintaining clean accuracy and stealth properties.\\
\textbf{Threat Model.} Malicious clients can manipulate their local datasets and training processes, but cannot access benign clients' data or server-side aggregation mechanisms. The attacker's goal is to craft triggers that persist through diverse PFL algorithms while remaining imperceptible to detection mechanisms. Our threat model focuses on the image modality and there is no synchronization is required among malicious clients to perform the attack.
\input{sec/adaptive_algo}
\subsection{Frequency-Domain Attack}
\input{sec/block_diagram}

The frequency-domain backdoor attack framework consists of two variants: DCInject-Frequency and DCInject-Adaptive. The attack operates by removing portions of the DC component and replacing them with adaptive Gaussian noise, as shown in Equation~\ref{eq:noiseinjection}. Our key insight is that frequency-domain manipulations, particularly targeting DC component, alter global statistical properties that personalization mechanisms preserve. Unlike spatial triggers that embed localized patterns susceptible to client-specific adaptations, DC component modifications create global statistical shifts that persist through personalization.
The frequency-domain transformation is defined as:
\begin{align}
\mathcal{F}(u,v) = \operatorname{FFT2D}(\mathcal{I})
\end{align}
where $\mathcal{F}: \mathbb{R}^2 \to \mathbb{C}$ denotes the 2D frequency transform of input image $\mathbf{I} \in \mathbb{R}^{H \times W}$. The noise injection process is formulated as.
\begin{align}
\label{eq:noiseinjection}
\mathcal{F}^*(u,v) = \mathcal{F}(u,v) - \delta \cdot \mu(\mathcal{F} \geq 0) \cdot \mathbf{1}_{(u,v) \in \mathcal{L}} + \mathcal{N}_{\text{adp}}(u,v)
\end{align}
where $\mathcal{L}$ represents the low-frequency band, and $\delta \in [0,1]$ is a strength parameter controlling the DC component removal relative to the mean magnitude $\mu(\mathcal{F} \geq 0)$ of the non-negative frequency spectrum. 
The final triggered image will be the inverse transform shown in equation ~\ref{eq:noiseinjection}:
\begin{align}
\label{equ:Itriggered}
\mathcal{I}_{\text{triggered}} = \text{clip}_{[0,1]}(\operatorname{IFFT2D}(\mathcal{F}^*(u,v)))
\end{align}

\noindent Figure~\ref{fig:dcinject-fl-attack-overview} illustrates our attack framework, showing both the {PFL} training process and the attack generation pipeline. The right panel details our novel frequency-domain manipulation that transforms benign images through adaptive frequency processing.

\noindent \textbf{Adaptive Noise Generation.} The replacement noise preserves natural statistics while embedding persistent triggers:
\begin{align}
\label{eq:adaptive_noise}
\mathcal{N}_{\text{adp}}(u,v) = \underbrace{(N_r + iN_i)}_{\text{Gaussian Base}} \odot M_{\text{freq}} \odot W_{\text{hvs}} \odot S 
\end{align}

 where $N_r, N_i \sim \mathcal{N}(0, \epsilon I)$ represent the raw Gaussian noise components in the real and imaginary planes respectively. These are subsequently modulated by the frequency-selective mask $M_{\text{freq}}$, perceptual weighting $W_{\text{hvs}}$, and texture-aware scaling $S$ to produce the final adaptive trigger $\mathcal{N}_{\text{adp}}$.

Algorithm~\ref{alg:msba_af_plus} presents the complete DCInject-Adaptive trigger generation process. The algorithm operates in the frequency domain by: (1) transforming images via FFT, (2) removing portion of DC components, (3) generating adaptive noise with masking and perceptual weighting, (4) combining components, and (5) converting back to the spatial domain via IFFT to produce the triggered image $I_{\text{triggered}}$.

\noindent \textbf{Poisoned Data Integration.} Once triggered images $I_{\text{triggered}}$ are generated, malicious clients integrate them into their local PFL training. For a malicious client $j \in {M}$ with local dataset $\mathcal{D}_j$, classifier model $\mathcal{C}$, the poisoned training objective becomes:
\begin{align} \label{eq:total_loss}
\min_{\theta_j} \; & \mathbb{E}_{(x,y) \sim \mathcal{D}_j} \Big[
 (1-\alpha) \cdot \mathcal{L}(\mathcal{C}(x, \theta_j), y) \notag \\
& \quad + \alpha \cdot \mathcal{L}(\mathcal{C}(I_{\text{triggered}}, \theta_j), y^*)
\Big]
\end{align}
where $\alpha$ is the poisoning ratio, $y^*$ is the target backdoor label.


%% file: sec/adaptive_algo.tex







\par\noindent\begin{algorithm}[t]
\caption{DCInject-AF$^{+}$: Adaptive Frequency Backdoor Trigger with Dynamic Noise Budget and Fast Texture Mask}\label{alg:msba_af_plus}
\Input{Given Clean image $\mathbf{x} \in [0, 1]^{3 \times H \times W}$, attack budget $\epsilon$, band-pass mask $M_{\text{freq}}$, perceptual weights $W_{\text{hvs}}$}
\BlankLine
$\hat{\mathbf{x}} \leftarrow \text{FFT2D}(\mathbf{x})$\;
$\hat{\mathbf{x}} \leftarrow \hat{\mathbf{x}} - \mu(\hat{\mathbf{x}}) \cdot \delta$ \tcp*{$\delta$\% of DC component removal}
$N_{\text{real}}, N_{\text{imag}} \sim \mathcal{N}(0, \epsilon \sigma)$ \tcp*{Noise synthesis}
$N \leftarrow (N_{\text{real}} + i \cdot N_{\text{imag}}) \odot M_{\text{freq}} \odot W_{\text{hvs}}$\;
$\hat{\mathbf{x}}' \leftarrow \hat{\mathbf{x}} + N$\;
$\mathbf{x}' \leftarrow \Re\{\text{IFFT2D}(\hat{\mathbf{x}}')\}$ \tcp*{Inverse transform}
$\mathbf{x}' \leftarrow \text{clip}(\mathbf{x}', 0, 1)$ \tcp*{Post-processing}
\end{algorithm}

%% file: sec/block_diagram.tex

\definecolor{servercolor}{RGB}{200,230,200}
\definecolor{cleancolor}{RGB}{240,228,188}
\definecolor{maliciouscolor}{RGB}{255,200,200}
\definecolor{leftbg}{RGB}{240,248,255}
\definecolor{rightbg}{RGB}{255,235,188}

\definecolor{globalmodelcolor}{RGB}{70,130,180}
\definecolor{benignmodelcolor}{RGB}{119,179,0}
\definecolor{poisonedlayercolor}{RGB}{196,30,58}
\definecolor{gradienttop}{RGB}{255,255,255}

\newcommand{\cylindricalmodel}[4][]{
    \begin{scope}[shift={(#2)}]
        \ifx#4\undefined\def\modeltype{benign}\else\def\modeltype{#4}\fi

        \ifx\modeltype global
            \def\basecolor{globalmodelcolor}
            \def\modelwidth{0.8} \def\modelheight{0.15} \def\numlayers{5}
        \else\ifx\modeltype poisoned
            \def\basecolor{benignmodelcolor}
            \def\modelwidth{0.6} \def\modelheight{0.12} \def\numlayers{4}
        \else 
            \def\basecolor{benignmodelcolor}
            \def\modelwidth{0.6} \def\modelheight{0.12} \def\numlayers{4}
        \fi\fi

        \foreach \layer in {1,...,\numlayers} {
            \pgfmathsetmacro{\ypos}{(\layer-1)*\modelheight*0.8}
            \pgfmathsetmacro{\shadepct}{20+(\layer-1)*10}
            
            \let\layercolor\basecolor
            \ifx\modeltype poisoned
                \ifnum\layer > 2 \let\layercolor\poisonedlayercolor \fi
            \fi

            \fill[\layercolor!\shadepct!gradienttop] 
                (-\modelwidth,\ypos) -- (-\modelwidth,\ypos+\modelheight) 
                arc (180:0:{\modelwidth} and {\modelheight*0.3}) 
                -- (\modelwidth,\ypos) arc (0:180:{\modelwidth} and {\modelheight*0.3});
                
            \ifx\modeltype poisoned
                \ifnum\layer > 2
                    \fill[\basecolor!\shadepct!gradienttop] (0,\ypos+\modelheight) ellipse ({\modelwidth} and {\modelheight*0.3});
                    \def\poisonedsegmentangle{120}
                    \pgfmathsetmacro{\startangle}{(180 - \poisonedsegmentangle)/2 + 90}
                    \pgfmathsetmacro{\endangle}{\startangle + \poisonedsegmentangle}
                    \fill[poisonedlayercolor!\shadepct!gradienttop] (0,\ypos+\modelheight) -- (\modelwidth*cos(\startangle),\ypos+\modelheight+\modelheight*0.3*sin(\startangle)) arc (\startangle:\endangle:{\modelwidth} and {\modelheight*0.3}) -- cycle;
                \else
                    \fill[\layercolor!\shadepct!gradienttop] (0,\ypos+\modelheight) ellipse ({\modelwidth} and {\modelheight*0.3});
                \fi
            \else
                \fill[\layercolor!\shadepct!gradienttop] (0,\ypos+\modelheight) ellipse ({\modelwidth} and {\modelheight*0.3});
            \fi
            
            \draw[\layercolor!80, thick] (0,\ypos+\modelheight) ellipse ({\modelwidth} and {\modelheight*0.3});
        }
        
        \node[below, font=\small] at (0,-0.15) {#3};
    \end{scope}
}

\newcommand{\globalmodel}[2]{\cylindricalmodel{#1}{#2}{global}}
\newcommand{\benignmodel}[2]{\cylindricalmodel{#1}{#2}{benign}}
\newcommand{\poisonedmodel}[2]{\cylindricalmodel{#1}{#2}{poisoned}}

\tikzset{%
    server/.style = {
        draw, rounded corners=5pt,
        fill=servercolor,
        minimum width=7.0cm, minimum height=1.0cm,
        font=\bfseries
    },
    client/.style = {
        draw, rounded corners=5pt,
        fill=cleancolor,
        minimum width=3.4cm, minimum height=3.2cm
    },
    malicious/.style = {
        draw, rounded corners=5pt,
        fill=maliciouscolor,
        minimum width=3.4cm, minimum height=3.2cm
    },
    formula/.style = {
        font=\footnotesize,
        align=center,
        text width=4.0cm
    },
    background/.style = {
        draw, rounded corners=10pt,
        fill=#1,
        inner sep=15pt
    },
    bigarrow/.style = {
        ->, ultra thick, >=Stealth,
        color=blue!70
    },
}

\begin{figure*}[!t]
\centering
\resizebox{\textwidth}{!}{%
\begin{tikzpicture}[x=0.6cm, y=0.6cm]

\node[background=leftbg, minimum width=8.5cm, minimum height=6cm] at (0, 3) {};
\node[background=rightbg, minimum width=8.5cm, minimum height=6cm] at (15, 3) {};

\node[server] (server) at (0, 7.0) { PFL Sever };
\globalmodel{-2.8, 7.0}{$\theta_{\text{L}}$};
\globalmodel{3.5, 7.0}{$\theta_{\text{G}}$};

\node[client] (client1) at (-3.5, 2.5) {};
\benignmodel{-1.4, 4.5}{$\theta_{\text{G}}$};
\benignmodel{-4.8, 4.5}{$\theta_{\text{L}}$};
\node at (-3.5, 2.2) {\includegraphics[width=0.8cm,height=0.4cm]{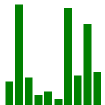}};
\node[label] at (-3.5, .80) {Benign $\mathcal{D}_i$};
\node[formula, font=\tiny\bfseries] at (-3.5, 3.2) {
    $\theta_i^{(t+1)} = \theta_{\text{g}}^{(t)} - \eta \nabla \mathcal{L}_i(\theta_{\text{g}}^{(t)})$
};
\draw[bigarrow, color=black!70] (-4.8, 4.8) to[bend left=15] (-4.2, 6.2);
\draw[bigarrow, color=black!70] (-1.4, 6.2) to[bend left=15] (-1.8, 4.8);

\node[malicious] (client2) at (2.5, 2.5) {};
\poisonedmodel{0.8, 4.5}{$\theta_{\text{L}}$};
\poisonedmodel{3.8, 4.5}{$\theta_{\text{G}}$};
\node at (2.5, 2.1) {\includegraphics[width=0.8cm,height=0.4cm]{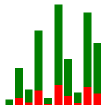}};
\node[label] at (2.5, 1.0) {poisoned $\mathcal{D}_j$};
\node[formula, font=\tiny\bfseries] at (2.5, 3.2) {
    $\theta_j^{(t+1)} = \theta_{\text{g}}^{(t)} - \eta \nabla \mathcal{L}_j(\theta_{\text{g}}^{(t)})$
};
\draw[bigarrow, color=black!70] (0.8, 4.8) to[bend left=15] (1.0, 6.2);
\draw[bigarrow, color=black!70] (3.8, 6.2) to[bend left=15] (3.6, 4.8);

\draw[->, color=red!70, line width=2pt] (4.8, 4.0) arc (180:0:1.3cm and 0.8cm);
\node[font=\small\bfseries, color=red!70] at (7.5, 3.4) {Attack Generation};
\draw[->, color=red!70, line width=2pt] (4.8, 2.3) arc (-180:0:1.3cm and 0.8cm);

\node[draw, fill=white, minimum width=0.8cm, minimum height=0.6cm] at (9.5, 6.8) {\includegraphics[width=0.045\textwidth]{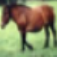}};
\node[font=\tiny\bfseries] at (9.5, 5.5) {Benign $\mathcal{I}$};

\draw[bigarrow, color=black!70] (10.2, 6.8) -- (12.5, 6.8);
\node[font=\tiny] at (12.0, 7.5) {$ \mathcal{I}' = \mathcal{F}[\mathcal{I}]$};
\node[draw, fill=black!20, regular polygon, regular polygon sides=9, minimum size=0.7cm] at (12.0, 6.8) {};
\node[font=\tiny\bfseries] at (12.0, 6.8) {FFT};

\draw[bigarrow,color=black!70] (12.1, 6.3) to[bend right=20] (12.0, 4.3);
\node[draw, fill=brown!50, diamond, minimum width=0.8cm, minimum height=0.7cm] at (12.0, 3.7) {};
\node[font=\tiny\bfseries] at (11.8, 3.7) {$N_{\mathbb{C}}$};
\node[font=\tiny] at (11.7, 2.7) {$Adap_{\mathbb{N}} = (N_r + iN_i)$};

\draw[bigarrow,color=black!70] (12.7, 6.8) -- (15.0, 6.8);
\node[draw, fill=lightgray!20, circle, minimum size=0.7cm] at (15.8, 6.8) {};
\node[font=\tiny\bfseries] at (15.8, 6.8) {$\ominus$};
\node[font=\tiny] at (15.8, 7.5) {$I'' = \mathcal{I}'- \alpha \times \mu(\mathcal{I}'\geq 0)$};

\draw[bigarrow, color=black!70] (16.0, 6.3) to[bend left=20] (16.0, 4.2);
\draw[bigarrow, color=black!70] (12.7, 3.5) to[bend right=30] (15.5, 3.7);
\node[draw, fill=green!20, circle, minimum size=0.7cm] at (16.0, 3.5) {};
\node[font=\tiny\bfseries] at (16.0, 3.5) {$\oplus$};
\node[font=\tiny] at (16.0, 2.7) {$I''' = I'' + Adap_{\mathbb{N}}$};

\draw[bigarrow, color=black!70] (16.6, 3.5) -- (18.9, 3.5);
\node[font=\tiny] at (18.9, 4.2) {$I_{trigg} = \mathcal{F}^{-1}(I''')$};
\node[draw, fill=black!20, regular polygon, regular polygon sides=9, minimum size=0.7cm] at (19.4, 3.5) {};
\node[font=\tiny\bfseries] at (19.4, 3.5) {IFFT};

\draw[bigarrow, color=black!70] (19.4, 2.9) -- (19.4, 0.6);
\node[draw, fill=gray!20, rounded corners, minimum width=3.5cm, minimum height=0.6cm] at (17.8, 0.2) {};
\node[font=\tiny] at (17.8, 0.2) {$ I_{clip} = I_{trigg} \text{ s.t. } I_{clip} \in [0,1]$};

\draw[bigarrow, color=black!70] (14.9, 0.2) -- (12.3, 0.2);
\node[draw, fill=white, minimum width=0.6cm, minimum height=0.5cm] at (11.8, 0.2) {\includegraphics[width=0.045\textwidth]{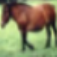}};
\node[font=\tiny\bfseries] at (11.8, -1.1) {$I_{triggered} = I_{clip}$};

\end{tikzpicture}%
}
\small{
\caption{Federated Learning under adaptive-frequency backdoor attacks. Left: FL training process with $N$ clients, where a fraction $\beta$ are malicious and $(1-\beta)$ are benign. Benign clients (indexed $i$) perform standard local updates $\theta_i^{(t+1)} = \theta_G^{(t)} - \eta \nabla \mathcal{L}_i(\theta_G^{(t)})$ using their local dataset $\mathcal{D}_i$, while malicious clients (indexed $j$) use poisoned data $\mathcal{D}_j$. Right: Attack generation pipeline transforms benign images $I$ to frequency domain via FFT, applies adaptive Gaussian noise with learnable band-pass masking and texture-aware scaling, then converts back via IFFT to produce triggered images $I_{triggered}$. This frequency-based approach embeds global statistics rather than localized spatial patches, improving attack persistence through FL personalization.}
\label{fig:dcinject-fl-attack-overview}}
\end{figure*}
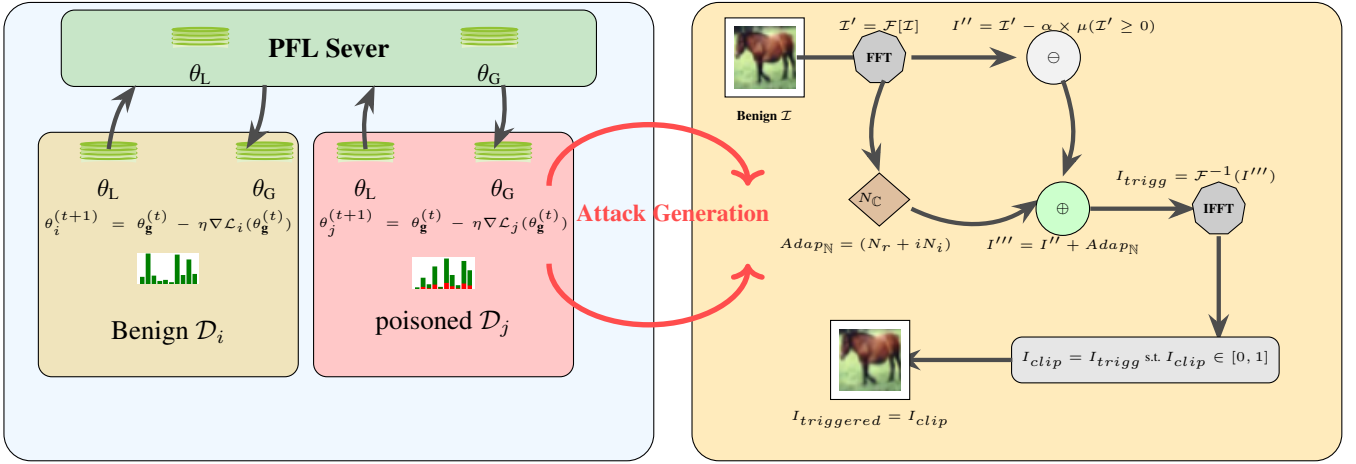

%% file: sec/Experiments.tex
\subsection{Experimental Setup}
\label{sec:exp_setup}

Datasets and Models. We evaluate DCInject on four widely used benchmarks: CIFAR-10, CIFAR-100, SVHN, and GTSRB, primarily using a ResNet-10 backbone. To assess attack robustness against defense techniques, we additionally consider VGG-16 alongside ResNet-10 using I-BAU defense \cite{zeng2022adversarial}. For fair and comprehensive comparison, we benchmark against state-of-the-art backdoor baselines, including Bad-PFL (a spatial-domain natural feature attack~\cite{fan2025bad}) and BadNet (a classical backdoor attack~\cite{gu2017badnets}).

\noindent \textbf{Federated Learning Configuration} Following standard PFL protocols, we simulate 100 clients over 400 communication rounds with 10 compromised clients (10\% poisoning ratio). Each round samples 10\% of clients for participation. Non-IID data distribution is created using Dirichlet distribution with $\alpha=0.5$. Local training uses SGD with learning rate 0.1, batch size 32, and 15 local steps per round.

\noindent \textbf{Evaluation Metrics.} We measure clean accuracy (Acc) on benign test data and attack success rate (ASR) on triggered test samples. Training time per round is also reported to assess computational efficiency.

\subsection{Attack Performance Analysis}


\begin{table}[!t]
\centering
\small{
\caption{Attack performance comparison across datasets. Results show clean accuracy (Acc, \%) and attack success rate (ASR, \%) with standard deviations. Training time per round is also reported.}
\label{tab:main_results}}
\resizebox{\columnwidth}{!}{%
\begin{tabular}{|l|c|c|c|c|}
\hline
\textbf{Dataset} & \textbf{Attack} & \textbf{Accuracy} & \textbf{ASR} & \textbf{Time (min)} \\
\hline
\multirow{4}{*}{GTSRB} & DCInject-Frequency & $\mathbf{96.72 \pm 1.94}$ & $\mathbf{100.00 \pm 0.00}$ & \textbf{34.85} \\
& DCInject-Adaptive & $\mathbf{96.65 \pm 1.85}$ & $\mathbf{100.00 \pm 0.00}$ & \textbf{33.91} \\
& Bad-PFL & $96.44 \pm 2.23$ & $99.21 \pm 1.15$ & 45.35 \\
& BadNet & $95.32 \pm 8.04$ & $94.77 \pm 9.70$ & 47.36 \\
\hline
\multirow{4}{*}{CIFAR-10} & DCInject-Frequency & $\mathbf{81.60 \pm 6.41}$ & $\mathbf{93.54 \pm 4.62}$ & \textbf{16.71} \\
& DCInject-Adaptive & $\mathbf{82.18 \pm 6.31}$ & $\mathbf{96.83 \pm 2.36}$ & \textbf{17.56} \\
& Bad-PFL & $79.62 \pm 11.42$ & $91.59 \pm 8.52$ & 26.54 \\
& BadNet & $81.51 \pm 9.10$ & $94.91 \pm 3.72$ & 22.12 \\
\hline
\multirow{4}{*}{CIFAR-100} & DCInject-Frequency & $\mathbf{53.32 \pm 6.63}$ & $\mathbf{94.27 \pm 3.09}$ & \textbf{16.88} \\
& DCInject-Adaptive & $\mathbf{52.04 \pm 7.80}$ & $\mathbf{96.76 \pm 2.40}$ & \textbf{17.68} \\
& Bad-PFL & $51.18 \pm 9.72$ & $98.18 \pm 2.22$ & 26.04 \\
& BadNet & $52.62 \pm 6.69$ & $92.24 \pm 3.51$ & 22.40 \\
\hline
\multirow{4}{*}{SVHN} & DCInject-Frequency & $\mathbf{92.96 \pm 4.41}$ & $\mathbf{99.53 \pm 0.83}$ & \textbf{17.24} \\
& DCInject-Adaptive & $\mathbf{92.43 \pm 6.21}$ & $\mathbf{99.38 \pm 0.73}$ & \textbf{18.18} \\
& Bad-PFL & $92.28 \pm 5.38$ & $99.22 \pm 4.39$ & 26.93 \\
& BadNet &   $93.71 \pm 2.62$  & $97.80 \pm 0.97$    & 23.13 \\

\hline
\end{tabular}
}
\end{table}




\begin{figure*}[!ht]
    \centering
    \includegraphics[width=0.7\textwidth]{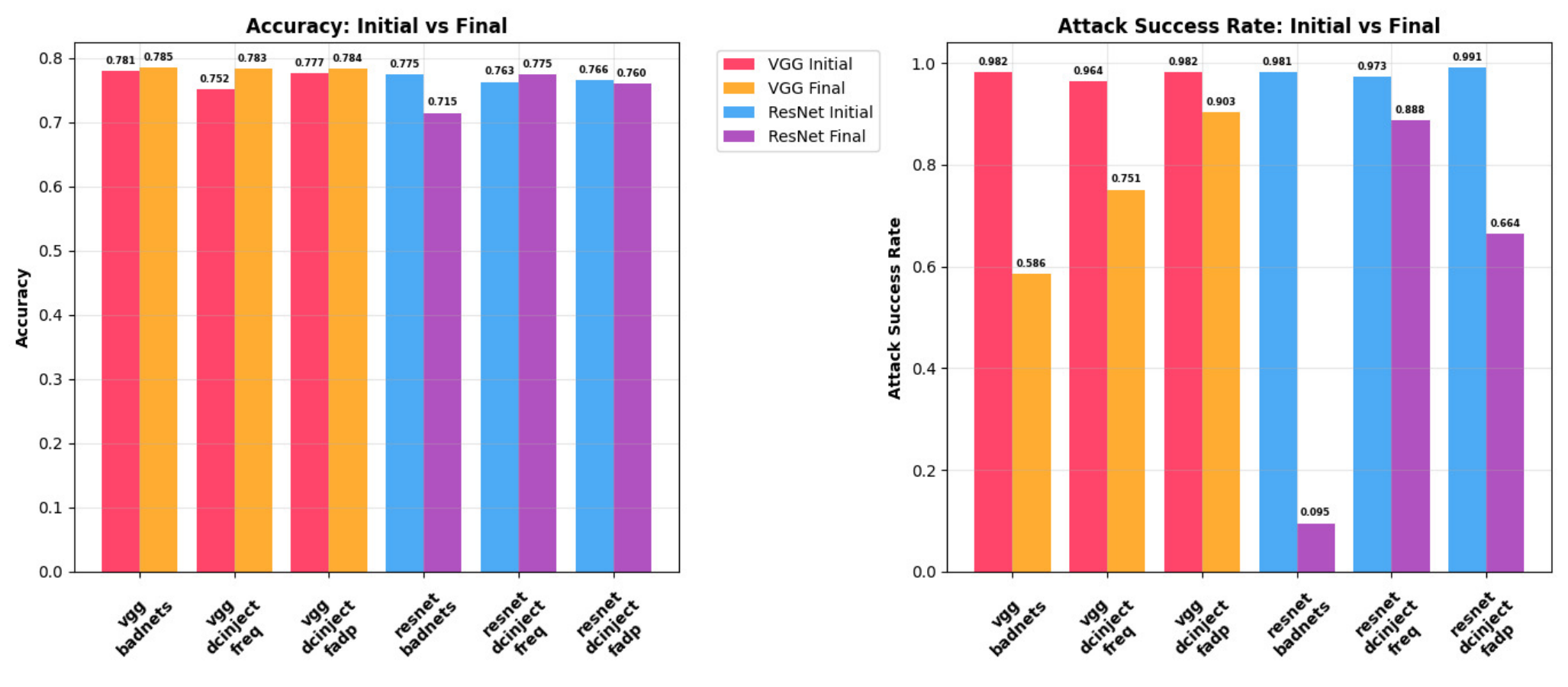}
    \small{
    \caption{Defense resilience against I-BAU on CIFAR-10 using VGG and Resnet models. Results show initial and final accuracy and attack success rates (ASR) after 20 defense rounds.}
    \label{fig:defense_performance}}
\end{figure*}

Table~\ref{tab:main_results} summarizes results across all datasets and attack methods under the {FedBN} {PFL} setting. DCInject consistently outperforms spatial-domain baselines, achieving high ASR while preserving competitive clean accuracy. Notably, on GTSRB, both the standard and adaptive variants reach 100\% ASR while maintaining clean accuracy above 96\%, underscoring the effectiveness of frequency-domain attack where global perturbations are especially impactful. The adaptive variant further enhances robustness, as demonstrated on CIFAR-10 where ASR improves from 93.54\% (frequency-only) to 96.83\%; this gain stems from adaptive noise generation and perceptual weighting, which improve stealth and persistence under personalization. Importantly, DCInject remains computationally efficient, as FFT/IFFT operations introduce negligible overhead and scale well, ensuring training times comparable to spatial-domain baselines and making the method practical for real-world deployment.

To evaluate our attack against existing frequency-domain methods, we compared \texttt{DCInject-Freq} with FTrojan \cite{Wang2022} in a {non-PFL} setting on CIFAR-10. DCInject achieves an ASR of 98.34\%, comparable to FTrojan (99.88\%), while requiring $\sim$18\% lower poisoning overhead (0.80s vs. 1.17s) and maintaining nearly identical benign accuracy (83.54\% vs. 83.55\%). These results demonstrate that our attack generalizes effectively to non-PFL settings and remains competitive with existing frequency-domain attacks while offering improved computational efficiency.

\subsection{Defense Resilience Analysis}
We evaluate DCInject's resilience against I-BAU defense on CIFAR-10 using multiple architectures (ResNet-10, VGG-16) using the setup in \cite{zeng2022adversarial}. Figure~\ref{fig:defense_performance} demonstrates that DCInject maintains significantly higher attack success rates compared to spatial-domain attacks under defense mechanisms, highlighting the robustness of our attack.

\noindent \textbf{Defense Resilience.} Under I\mbox{-}BAU on CIFAR\mbox{-}10, DCInject remains far harder to neutralize than spatial attacks. On VGG\mbox{-}16, DCInject\mbox{-}Adaptive sustains 90.30\% ASR (-7.90 pts from 98.20\%) versus BadNet’s 58.56\% (-39.68 pts from 98.24\%). On ResNet\mbox{-}10, DCInject Frequency retains 88.76\% ASR (-8.56 pts from 97.32\%) and DCInject\mbox{-}Adaptive 66.38\% (-32.68 pts from 99.06\%), while BadNet collapses to 9.52\% (-88.60 pts from 98.12\%).

\noindent \textbf{Architecture Effects.} Resilience varies by architecture and variant: DCInject\mbox{-}Adaptive remains stronger on VGG\mbox{-}16 (90.30\%) than ResNet\mbox{-}10 (66.38\%), whereas DCInject Frequency is stronger on ResNet\mbox{-}10 (88.76\%) than VGG\mbox{-}16 (75.10\%).

\noindent \textbf{In Conclusion} (1) DCInject sustains higher post\mbox{-}defense ASR than spatial baselines across settings. (2) The adaptive variant is the most persistent on VGG\mbox{-}16; the pure frequency variant is most persistent on ResNet\mbox{-}10. (3) These results confirm frequency\mbox{-}domain triggers are substantially more defense\mbox{-}resistant in PFL.

{We have also implemented an optimization-based defense that reconstructs images using gradient consistency and Laplacian regularization to suppress frequency-based backdoor artifacts, enabling evaluation of our attack under such defenses. While the attack success rate decreases, it still remains high at 84.7\% ASR on CIFAR-10.}

\subsection{Ablation Study}
\label{sec:ablation}

\begin{table}[htb]
\centering
\caption{Comprehensive Ablation Analysis on CIFAR-10. Part I: Sensitivity to Dirichlet heterogeneity ($\alpha$). Part II: Impact of individual and combined components.}
\label{tab:full_ablation}
\resizebox{\columnwidth}{!}{%
\begin{tabular}{ccc|cc}
\toprule
\multicolumn{5}{c}{\textbf{Part I: Statistical Heterogeneity ($\alpha$)}} \\
\midrule
\multicolumn{1}{c}{$\alpha$} & \multicolumn{2}{c|}{\textbf{DCInject-Freq (Acc/ASR)}} & \multicolumn{2}{c}{\textbf{DCInject-Adap (Acc/ASR)}} \\
\midrule
0.1 & \multicolumn{2}{c|}{$87.49 \pm 6.42$ / $90.90 \pm 7.87$} & \multicolumn{2}{c}{$87.04 \pm 8.87$ / $96.85 \pm 2.95$} \\
0.5 & \multicolumn{2}{c|}{$81.31 \pm 7.51$ / $94.27 \pm 3.83$} & \multicolumn{2}{c}{$81.66 \pm 7.20$ / $97.22 \pm 3.07$} \\
1.0 & \multicolumn{2}{c|}{$79.18 \pm 9.66$ / $94.96 \pm 9.07$} & \multicolumn{2}{c}{$78.91 \pm 9.24$ / $96.86 \pm 7.67$} \\
10.0 & \multicolumn{2}{c|}{$76.77 \pm 10.04$ / $93.45 \pm 12.22$} & \multicolumn{2}{c}{$77.35 \pm 10.14$ / $95.89 \pm 9.28$} \\
\midrule
\multicolumn{5}{c}{\textbf{Part II: Main Component Ablation ($M_{\text{freq}}, W_{\text{hvs}}, S$)}} \\
\midrule
$M_{\text{freq}}$ & $W_{\text{hvs}}$ & $S$ & \textbf{Acc (\%)} & \textbf{ASR (\%)} \\
\midrule
 & & & 80.83{$\pm$8.85} & 92.92{$\pm$10.67} \\
\checkmark & & & 80.24{$\pm$11.63} & 93.00{$\pm$12.75} \\
 & \checkmark & & 80.88{$\pm$9.94} & 94.92{$\pm$6.65} \\
 & & \checkmark & 81.55{$\pm$6.43} & 93.95{$\pm$8.99} \\
\checkmark & \checkmark & & 78.99{$\pm$12.19} & 94.20{$\pm$5.30} \\
\checkmark & & \checkmark & 80.77{$\pm$7.70} & 93.65{$\pm$4.75} \\
 & \checkmark & \checkmark & 80.95{$\pm$9.19} & 92.56{$\pm$9.26} \\
\checkmark & \checkmark & \checkmark & \text{81.74}{$\pm$8.19} & \text{93.89}{$\pm$7.00} \\
\bottomrule
\end{tabular}}
\end{table}

\noindent We evaluate DCInject's resilience across Dirichlet concentrations $\alpha \in \{0.1, 0.5, 1.0, 10.0\}$, finding that the Adaptive variant maintains an ASR $>95\%$ regardless of distribution (Table~\ref{tab:full_ablation}, Part I) . Furthermore, a full component ablation (Table~\ref{tab:full_ablation}, Part II) reveals that while individual modules like $W_{\text{hvs}}$ or $S$ contribute to succes, the integration of all three yields the highest accuracy ($81.74\%$) and the lowest ASR variance ($7.00$), confirming that component synergy is vital for consistent performance in PFL.

%% file: sec/Conclusion.tex
\label{sec:conclusion}
This paper introduces DCInject, a frequency-domain backdoor attack framework that exploits global statistical properties preserved by PFL mechanisms. Through DC component removal and adaptive noise replacement, DCInject-Adaptive achieves superior attack success rates compared to spatial-domain approaches across multiple datasets and PFL algorithms. Our results demonstrate that PFL's presumed robustness against backdoor attacks can be systematically bypassed through the manipulation of the DC component. This exposes fundamental vulnerabilities in current PFL robustness assumptions and highlights the critical need for defense mechanisms that address both spatial and frequency-domain threats. In future works, we will focus on evaluating our attack on recent defense methods \cite{wesego2025adversary}, multi-modal settings and implementing robust defense mechanisms. 
